\documentclass[twocolumn, prl, superscriptaddress, 10pt]{revtex4-1}
\usepackage{graphicx, subfigure, amsfonts,amssymb,amsmath, array, gensymb,fontenc,color, lipsum}

\begin{document}

\title{{Edge Disorder in Bottom-Up Zigzag Graphene Nanoribbons}: \\ Implications for Magnetism and Quantum Electronic Transport}

\author{Michele Pizzochero}
\email{mpizzochero@g.harvard.edu}
\affiliation{Institute of Physics, Ecole Polytechnique F\'ed\'erale de Lausanne (EPFL), CH-1015 Lausanne, Switzerland}
\affiliation{School of Engineering and Applied Sciences, Harvard University, Cambridge, Massachusetts 02138, United States}
\affiliation{National Centre for Computational Design and Discovery of Novel Materials (MARVEL), Ecole Polytechnique F\'ed\'erale de Lausanne (EPFL), CH-1015 Lausanne, Switzerland}

\author{Gabriela Borin Barin}
\affiliation{Nanotech@Surfaces Laboratory, Swiss Federal Laboratories for Materials Science and Technology (EMPA), CH-8600 D\"ubendorf, Switzerland}
\affiliation{These authors contributed equally to this work}

\author{Kristi\={a}ns \v{C}er\c{n}evi\v{c}s}
\affiliation{Institute of Physics, Ecole Polytechnique F\'ed\'erale de Lausanne (EPFL), CH-1015 Lausanne, Switzerland}
\affiliation{National Centre for Computational Design and Discovery of Novel Materials (MARVEL), Ecole Polytechnique F\'ed\'erale de Lausanne (EPFL), CH-1015 Lausanne, Switzerland}

\author{Shiyong Wang}
\affiliation{Nanotech@Surfaces Laboratory, Swiss Federal Laboratories for Materials Science and Technology (EMPA), CH-8600 D\"ubendorf, Switzerland}

\author{Pascal Ruffieux}
\affiliation{Nanotech@Surfaces Laboratory, Swiss Federal Laboratories for Materials Science and Technology (EMPA), CH-8600 D\"ubendorf, Switzerland}

\author{Roman Fasel}
\affiliation{Nanotech@Surfaces Laboratory, Swiss Federal Laboratories for Materials Science and Technology (EMPA), CH-8600 D\"ubendorf, Switzerland}
\affiliation{Department of Chemistry, Biochemistry and Pharmaceutical Sciences, University of Bern, CH-3012 Bern, Switzerland}

\author{Oleg V. Yazyev}
\affiliation{Institute of Physics, Ecole Polytechnique F\'ed\'erale de Lausanne (EPFL), CH-1015 Lausanne, Switzerland}
\affiliation{National Centre for Computational Design and Discovery of Novel Materials (MARVEL), Ecole Polytechnique F\'ed\'erale de Lausanne (EPFL), CH-1015 Lausanne, Switzerland}

\begin{abstract}
We unveil the nature of the structural disorder in bottom-up zigzag graphene nanoribbons along with its effect on the magnetism and electronic transport on the basis of scanning probe microscopies and first-principles calculations. We find that edge-missing $m$-xylene units emerging during the cyclodehydrogenation step of the on-surface synthesis are the most common point defects. These  ``bite''  defects act as spin-1 paramagnetic centers, severely disrupt the conductance spectrum around the band extrema, and give rise to spin-polarized charge transport. We further show that the electronic conductance across graphene nanoribbons is more sensitive to ``bite'' defects forming at the zigzag edges than at the armchair ones. Our work establishes a comprehensive understanding of the low-energy electronic properties of disordered bottom-up graphene nanoribbons.
\end{abstract}

\maketitle

Graphene nanoribbons (GNRs), few-atom wide strips of graphene, are rapidly emerging as promising building blocks for next-generation nanotechnologies by virtue of the pronounced interplay between their crystal and electronic structure \cite{Yaz13}. GNRs can be fabricated in an atomically precise fashion by resorting to a bottom-up route that consists of surface-assisted Ullman-type coupling of organic precursor monomers and subsequent cyclodehydrogenation \cite{Cai10a}. Depending on the molecular structure of the monomer, full control over the target products is unambiguously achieved \cite{Cai10a, Tarliz16}, hence enabling the synthesis of a broad range of graphene nanoribbons with well-defined edge geometries \cite{Talirz17, Ruffieux2016, Nguy17, Liu2015, Groning2018}, widths \cite{Chen13, Chen15a, Jacobse2017, Wang2017, Pizzochero2020b, Pizzochero2021}, and heteroatom incorporations \cite{Nguyen2016, Friedrich2020}.

Among the wide spectrum of atomically precise graphene edges that have been manufactured via on-surface synthesis \cite{Yano20}, zigzag graphene nanoribbons (ZGNRs) are unique owing to their unconventional metal-free magnetic order \cite{Ruffieux2016, Li2019, Yazyev2010} that is predicted to be preserved up to room temperature \cite{Magda2014}. Irrespective of the width, ZGNRs possess magnetic moments that are coupled ferromagnetically along the edge and antiferromagnetically across it \cite{Fujita1996, Yazyev2010}. It has been shown that the electronic and magnetic structures can be modulated to a large extent via, e.g., charge doping \cite{Jung2010}, electric fields \cite{Son2006}, lattice deformations \cite{Zhang2017}, or defect engineering \cite{Wimmer2008, Ortiz2016}. The combination of tunable magnetic correlations, sizable band-gap width \cite{Son06a}, and weak spin-orbit interactions \cite{Avsar2020} makes ZGNRs promising candidates for spin logic operations \cite{Yazyev2008}.

Although atomic-scale imperfections in graphene-based materials have been extensively studied owing to their central role in shaping a number of functionalities of the hosting crystal \cite{Hashimoto2004, Banhart2011, Nair2013, Gonzalez-Herrero437, Bonfanti2018}, their experimental observation in bottom-up nanoribbons remains very limited \cite{Costa2018, Pizzochero2020}. In this Letter, we address for the first time the emergence of point defects {in on-surface synthesized 6-carbon zigzag
lines wide graphene nanoribbons (6-ZGNRs)}. We identify ``bite'' defects and assess their underlying impact on the magnetism and quantum electronic transport in these nanoribbons. We additionally draw a systematic comparison between ``bite'' defects in zigzag and armchair graphene nanoribbons (AGNRs) {of equal width} to reach a wide-angle view of the structural disorder that invariably arises in their on-surface synthesis.

 \begin{figure}[t!]
  \centering
 \includegraphics[width=1\columnwidth]{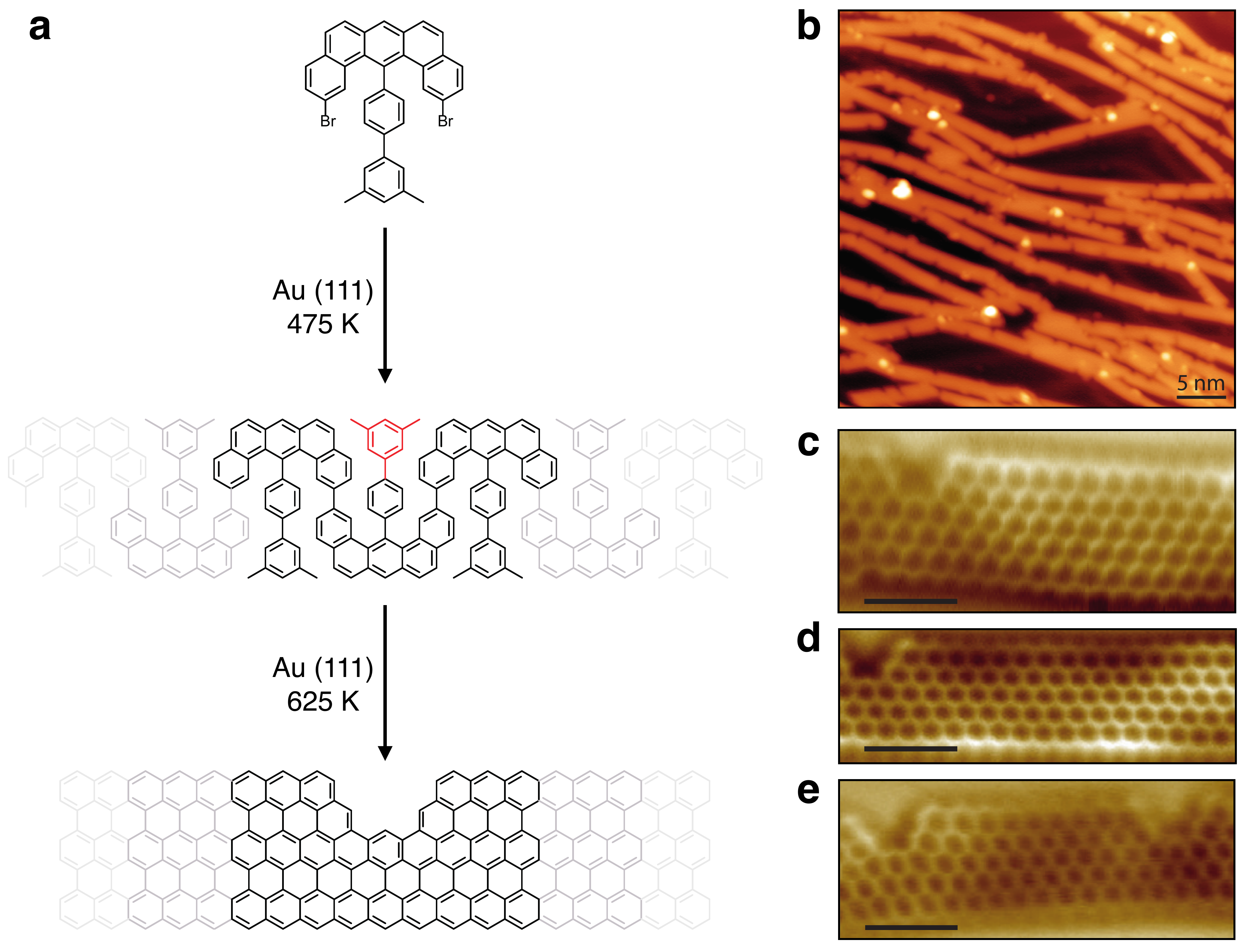}
  \caption{(a) Reaction scheme for the on-surface synthesis of 6-ZGNRs. The $m$-xylene unit that yields the formation of the ``bite'' defect in the final product is shown in red color. (b) STM topography image of 6-ZGNRs (LHe, 1 V, 0.01 nA). Notice the presence of missing carbon atoms at the edges. {The bright spots correspond to protrusions originating in highly disordered areas of the samples}. Constant height NC-AFM image taken with a CO-functionalized tip (resonant frequency 23.5 KHz, amplitude 70 pm) of (c, d) single and (e) double ``bite'' defects. Scale bars correspond to 1 nm. \label{Fig1}}
\end{figure}

{We focus on 6-ZGNR as, to the best of our knowledge, this is the only width of ZGNRs  that has been achieved via on-surface synthesis so far. \cite{Ruffieux2016}} The synthesis is carried out in ultra-high vacuum conditions employing 2,12-dibromo-14-(3',5'-dimethyl-[1,1'-biphenyl]-4-yl)-dibenzo[a,j]anthracene as the precursor monomer. The reaction scheme is given in Fig.\ \ref{Fig1}(a). The Au(111) (MaTeck GmbH) surface is initially cleaned by repeated cycles of Ar$^{+}$ sputtering at 1 keV and annealing at 470 $\textsuperscript{o}$C. Next, we thermally evaporate the precursor molecules on the clean surface, followed by a two-step annealing process, i.e., first at 475 K to activate the polymerization reaction, and then at 625 K to form the final 6-ZGNR product. 

In Fig.\ \ref{Fig1}(b), we show a representative scanning tunneling microscopy (STM) image that overviews our as-synthesized 6-ZGNRs on Au(111). A closer inspection reveals the presence of ubiquitous carbon vacancy defects located at the edges of the nanoribbons. 
We resolve their atomic structure by non-contact atomic-force microscopy (NC-AFM). Our results are shown in Fig.\ \ref{Fig1}(c-e) and indicate that each vacancy comprises a missing $m$-xylene unit, that is, a benzene ring with two methyl substituents attached in the relative meta position.  We dub this type of structural imperfection as ``bite'' defect. These defects originate from the scission of the C-C bond that occurs during the cyclodehydrogenation process of the reaction, as schematized in Fig.\ \ref{Fig1}(a). {Analogously to armchair-edged \cite{Pizzochero2020} and chevron-edged \cite{Costa2018} graphene nanoribbons, the atomic structure of the defect is dictated by the nature of the precursor monomer adopted in the on-surface synthesis}. By counting the number of defects in 150 distinct nanoribbons, we estimate the density of ``bite'' defects in our 6-ZGNRs samples to 0.26 $\pm$  0.10 nm$^{-1}$. This value is larger than that of equivalent defects in bottom-up AGNRs, as we show in Supplementary Fig.\ S1. \cite{Pizzochero2020}

 \begin{figure*}[t!]
  \centering
 \includegraphics[width=1.5\columnwidth]{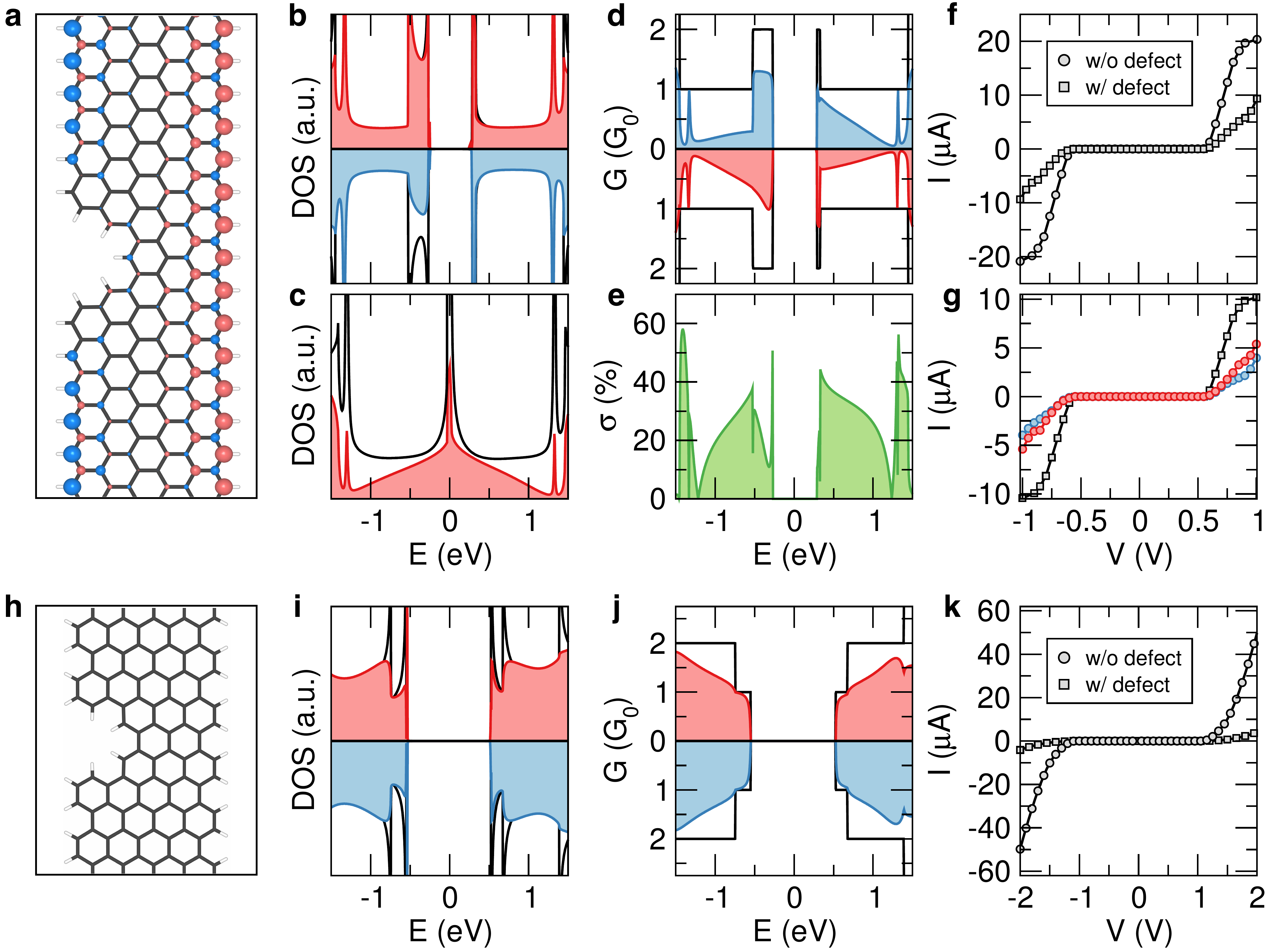}
  \caption{(a) Atomic structure and spin density of 6-ZGNR hosting a single ``bite'' defect. Red and blue colors indicate spin-majority and spin-minority channels, respectively. Electronic density of states of 6-ZGNR with (colored lines) and without (black lines) a ``bite'' defect in the (b) antiferromagnetic and (c) non-magnetic phases. (d) Zero-bias conductance spectrum of 6-ZGNR with (colored lines) and without (black lines) a ``bite'' defect. (e) Spin-polarization of the zero-bias conductance in 6-ZGNR hosting a ``bite'' defect. (f) $I$-$V$ curve of 6-ZGNR with (squares) and without (circles) a ``bite'' defect, along with (g) the contribution of spin-majority (red) and spin-minority (blue) currents. (h) Atomic structure of a single ``bite'' defect in 10-AGNR. (i) Electronic density of states and (j) zero-bias conductance spectrum of 10-AGNR with (colored lines) and without (black lines) a ``bite'' defects. (k) $I$-$V$ curve of 10-AGNR with (squares) and without (circle) a ``bite'' defect. \label{Fig2}}
\end{figure*}

Next, we theoretically address the effect of ``bite'' defects imaged in Fig.\ \ref{Fig1}(c,d) on the electronic structure and quantum transport properties of 6-ZGNRs. We rely on a combination of spin-polarized density-functional theory under the gradient-corrected approximation devised by Perdew, Burke, and Ernzerhof \cite{PBE} with non-equilibrium Green's functions. We take advantage of the \textsc{siesta} \cite{SIESTA} and \textsc{transiesta} \cite{TRANSIESTA} packages. Further details of our computations are given in Supplementary Note S1. The formation of the ``bite'' defect introduces an imbalance in the bipartite lattice that stems from the removal of five carbon atoms from sublattice $A$ and three atoms from sublattice $B$. In agreement with Lieb's theorem for the repulsive Hubbard model of a bipartite lattice at half-filling \cite{Lieb1989}, this sublattice-symmetry breaking translates to a net magnetic moment $m =2$ $\mu\textsubscript{B}$. Our results indicate that each ``bite'' defect acts as a $S = 1$ paramagnetic center in otherwise antiferromagnetic ($S = 0$) 6-ZGNRs. In Fig.\ \ref{Fig2}(a), we display the spin density in the vicinity of the defective site, which is found to be quenched at the lattice positions that are the closest to the missing atoms and to be fully restored $\sim$0.7 nm far from the ``bite'' defect. The defect does not lead to localized in-gap states and its effect on the density of states is mainly restricted to a reduction of the spin-minority channel, see Fig.\ \ref{Fig2}(b). As shown in Fig.\ \ref{Fig2}(c), such a reduction is spread over a quite ample energy window when the non-magnetic phase of 6-ZGNR is considered.

In Fig.\ \ref{Fig2}(d), we compare the zero-bias conductance spectrum of a 6-ZGNR with and without the ``bite'' defect.  The introduction of the defect causes a significant disruption of the conductance within both the valence band maximum (VBM) and conduction band minimum (CBM). We quantify this effect through a dimensionless descriptor, $\tau$, which accounts for the fraction of the conductance that is retained in the vicinity of the band extrema (here taken to be $\delta E = 0.10$ eV) in defective 6-ZGNR as \cite{Pizzochero2020}
\begin{equation}
\tau = \frac{\int_{\textnormal{VBM}-\delta E}^{\textnormal{CBM} +\delta E} G\textsubscript{d}(E) dE}{\int_{\textnormal{VBM}-\delta E}^{\textnormal{CBM} +\delta E} G\textsubscript{p}(E) dE}  \%,
\end{equation}

\noindent
where $G\textsubscript{d}(E)$ and $G\textsubscript{p}(E)$ is the zero-bias conductance spectrum pertaining to the defective and pristine graphene nanoribbons, respectively.  We obtain $\tau =$ 54\%, implying a substantial suppression of the conductance at the band edges ascribed to the defect. The defect-induced local magnetic moment gives rise to a net spin-polarization of the conductance, $\sigma (E)$, that is customarily ascertained as

\begin{equation}
\sigma(E) = \left |\frac{G\textsubscript{d}{^\uparrow}(E) - G\textsubscript{d}{^\downarrow}(E)}{G\textsubscript{d}{^\uparrow}(E) + G\textsubscript{d}{^\downarrow}(E)}  \right | \%,
\end{equation}

\noindent
with $G\textsubscript{d}{^\uparrow}(E)$ [$G\textsubscript{d}{^\downarrow}(E)$] being the zero-bias conductance spectrum of the spin-majority [spin-minority] channel of the defective graphene nanoribbon. In Fig.\ \ref{Fig2}(e), we show that $\sigma(E)$ attains values larger than 40\% at the band edges, demonstrating that even a low density of ``bite'' defects can promote a substantial spin-polarization of the charge carriers in 6-ZGNR. The detrimental impact of the ``bite'' defect on the electronic transport is reflected in the $I$-$V$ characteristics given in Fig.\ \ref{Fig2}(f). The current intensities that arise when the applied bias voltage exceeds in magnitude the band-gap width of 6-ZGNR are approximately halved in the defective graphene nanoribbon as compared to the pristine one. The local magnetic moment associated with the ``bite'' defect yields the formation of appreciable spin-polarized currents, as we show in Fig.\ \ref{Fig2}(g). Overall, our findings suggest that, though adverse to the performance of electronic devices, ``bite'' defects in 6-ZGNRs and their accompanying spin-1 magnetism can be exploited for spin-injection applications in carbon-based spintronics \cite{Wimmer2008}. 

We extend our investigation by comparing the influence of ``bite'' defects on the electronic transport across zigzag- and armchair-edged graphene nanoribbons. Given that the width of the hosting nanoribbons was demonstrated to be instrumental in governing the electronic transport properties, with AGNRs having a multiple-of-three width being subject to the largest defect-induced disruption of the conductance,  in the following we consider 10-AGNR, as the width of this nanoribbon (1.11 nm) is the closest to that of 6-ZGNR (1.14 nm). While the ``bite'' defect in the zigzag nanoribbon consists of a missing $m$-xylene unit (eight missing carbon atoms), earlier experimental works indicated that in the armchair congeners they consist of a missing benzene ring (six missing carbon atoms) \cite{Talirz17, Pizzochero2020}, see Fig.\ \ref{Fig2}(h). This difference in the atomic structure of the ``bite'' defect traces back to the necessarily different precursor monomer that is employed in their on-surface synthesis  \cite{Talirz17}. The formation of the ``bite'' defect in armchair graphene nanoribbons retains the sublattice symmetry, thus preserving the non-magnetic character of the hosting nanoribbon \cite{Lieb1989}. Similarly to the 6-ZGNR, the change that the ``bite'' defect promotes on the electronic density of states of 10-AGNR is limited to a depletion around the band extrema and again no in-gap states emerge in Fig.\ \ref{Fig2}(i). Contrary to 6-ZGNR, however, only a mild disruption in the vicinity of the band edges of 10-AGNR occurs in the zero-bias conductance spectrum given in Fig.\ \ref{Fig2}(j). The corresponding $\tau$ is found to be 77\%, a value that is 23\% larger than that of defective zigzag nanoribbons of equal width. In Fig.\ \ref{Fig2}(k), we show that, upon the application of finite bias voltages, a drastic reduction of the current intensity associated with the defect is observed in 10-AGNR.

 \begin{figure*}[t!]
  \centering
 \includegraphics[width=1.8\columnwidth]{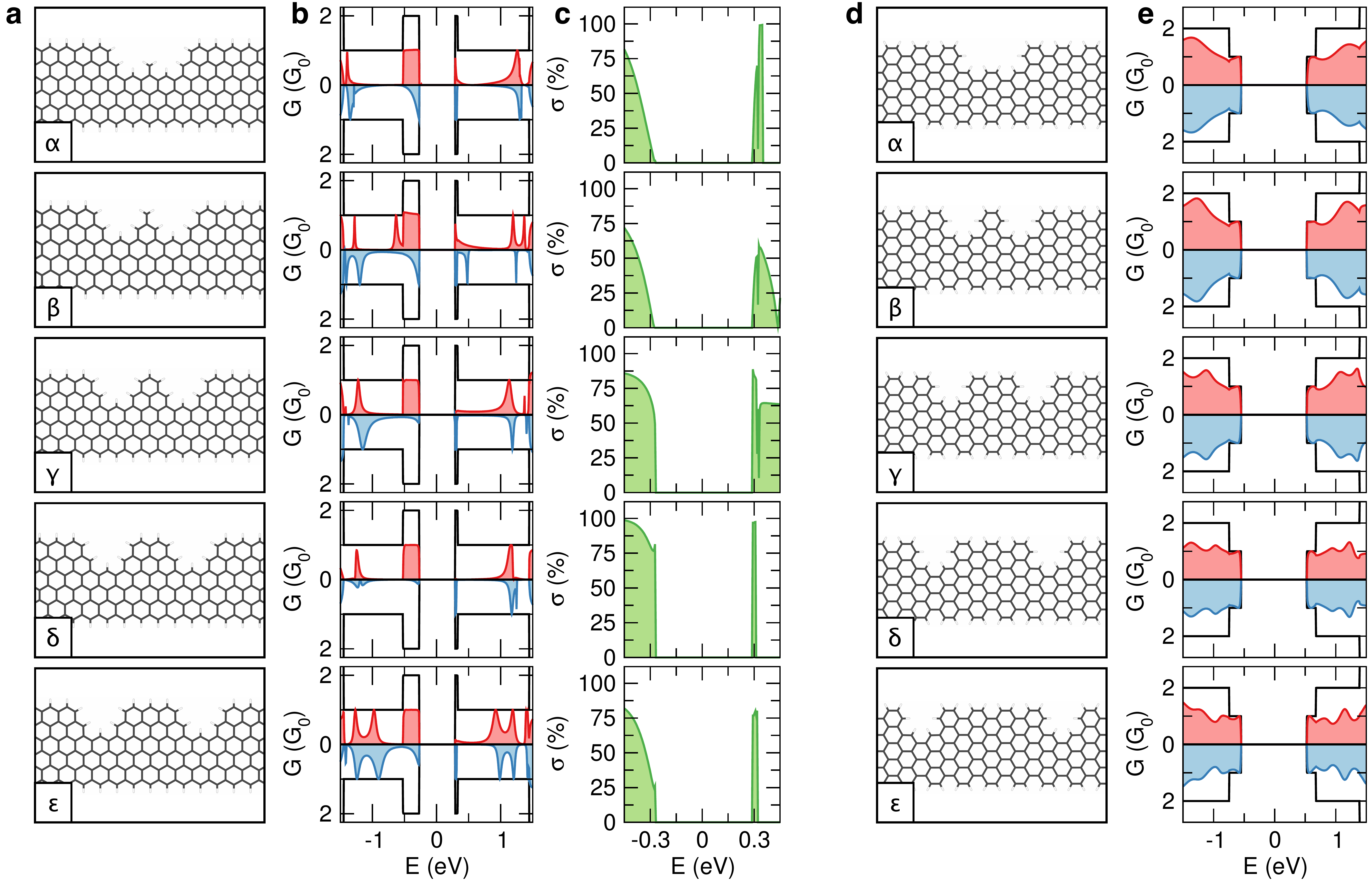}
  \caption{(a) Atomic structures of 6-ZGNR containing pairs of ``bite'' defects located at an increasing relative distance, along with the corresponding (b) zero-bias conductance spectrum and (c) spin-polarization of the conductance. (d) Atomic structures of 10-AGNR containing pairs of ``bite'' defects located at an increasing relative distance, along with the corresponding (e) zero-bias conductance spectrum.   \label{Fig3}}
\end{figure*}

Our non-contact atomic force microscopy imaging reveals an occasional aggregation of ``bite'' defects in 6-ZGNRs, see Fig.\ \ref{Fig1}(e). We generalize this experimental observation by modeling a pair of ``bite'' defects located at increasing relative distances (up to $\sim$1.5 nm), giving rise to the $\alpha$, $\beta$, $\gamma$, $\delta$ and $\epsilon$ configurations that are displayed in Fig.\ \ref{Fig3}(a). As is evident from Fig.\ \ref{Fig3}(b), pairs of ``bite'' defects further diminish the zero-bias conductance at the band edges, with $\tau$ attaining values of 32\%, 35\%, 22\%, 19\%, 25\% for the $\alpha$, $\beta$, $\gamma$, $\delta$ and $\epsilon$ configurations, respectively. This indicates that the conductance around the band extrema is halved as compared to the introduction of a single ``bite'' defect, for which $\tau =$ 54\%, and that multiple ``bite'' defects severely hinder electronic transport in zigzag nanoribbons.  The formation of the secondary defect further affects the sublattice- and spin-imbalance by increasing the magnetic moment from 2 $\mu\textsubscript{B}$ to 4 $\mu\textsubscript{B}$ for all the configurations of defect pairs considered but the $\alpha$ one, for which it increases to 3 $\mu\textsubscript{B}$. This, in turn, consistently  enhances the spin-polarization of the conductance presented in Fig.\ \ref{Fig3}(c), which reaches unitary values in, e.g.,  the conduction states of the $\alpha$ and $\delta$ defect configurations. ({Full spin polarization at the band extrema can be achieved by periodically arranging a high density of "bite" defects in 6-ZGNRs, as we show in Supplementary Fig.\ S2}). A different scenario emerges in armchair-edged nanoribbons when pairs of ``bite'' defects are considered in the arrangements shown in Fig.\ \ref{Fig3}(d). The introduction of a secondary defect in 10-AGNR has a very minor effect around the band edges in the zero-bias conductance spectra shown in Fig.\ \ref{Fig3}(e). Surprisingly, $\tau$ is found to increase from 77\% for the single ``bite'' defect to $\sim$90\%, irrespective of the relative distance between of defect pairs. Overall, these findings suggest that the electronic transport across 10-AGNR is more robust to ``bite'' defects as compared to 6-ZGNRs.

In summary, we have identified ``bite'' defects, i.e., missing $m$-xylene units at the edges, as the dominant source of atomic-scale disorder in bottom-up {6-carbon zigzag lines wide} graphene nanoribbons. The formation of these unintentional defects occurs during the cyclodehydrogenation step of the on-surface synthesis and  induces sublattice- and spin-imbalance, thus causing a nominal local magnetic moment of 2 $\mu \textsubscript{B}$. This, in turn, gives rise to spin-polarized charge transport that renders defective zigzag nanoribbons potential platforms for applications in all-carbon logic spintronics in the ultimate limit of scalability. We have additionally conducted a systematic comparison between the zigzag-edged and armchair-edged nanoribbons of equal width, finding that the transport across the former is less sensitive than the latter upon the introduction of both single and multiple defects. To conclude, our work offers a global picture of the impact of ubiquitous ``bite'' defects on the low-energy electronic structure of bottom-up graphene nanoribbons. {Potential future research directions include the investigation of other types of point defects experimentally observed at the edges of on-surface synthesized nanoribbons \cite{Ruffieux2016, Costa2018}.}

\smallskip
\noindent
M.P.\ and G.\ B.\ B.\ contributed equally to this work. M.P., K.\v{C}., and O.V.Y. are financially supported by the Swiss National Science Foundation under grant No.\ 172543 and NCCR MARVEL. M.P.\ is partly supported by the Swiss National Science Foundation through the Early Postdoc.Mobility program (Grant No.\ P2ELP2-191706). G.B.B., S.W., P.R., and R.F. acknowledge funding from the Swiss National Science Foundation under the grant No.\ 200020\_182015 and IZLCZ2\_170184, the European Union Horizon 2020 research and innovation program under the grant No.\ 881603 (Graphene Flagship Core 3), and the Office of Naval Research (N00014-18-1-2708). First-principles calculations have been performed at the Swiss National Supercomputing Center (CSCS) under project s1008.

\providecommand{\latin}[1]{#1}
\providecommand*\mcitethebibliography{\thebibliography}
\csname @ifundefined\endcsname{endmcitethebibliography}
  {\let\endmcitethebibliography\endthebibliography}{}

\end{document}